\documentclass{moriond}

\def\Journal#1#2#3#4{{#1} {\bf #2}, #3 (#4)}

\def\PRD{{\em Phys. Rev.} D}

\def\be{\begin{equation}}
\def\ee{\end{equation}}
\def\bea{\begin{eqnarray}}
\def\eea{\end{eqnarray}}

\def\pt{$p_{\mathrm{T}}$}

\newcommand{\Photo}{\includegraphics[height=34mm]{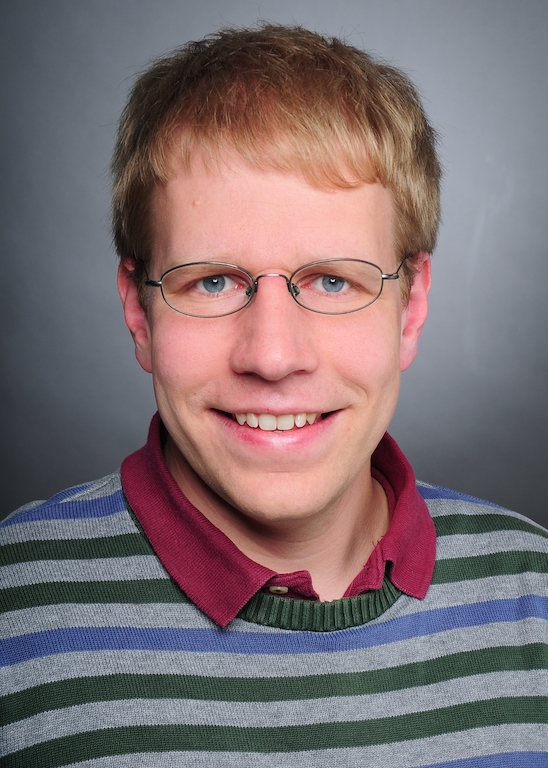}}

\begin{document}
\vspace*{4cm}
\title{Standard Model QCD with jets and photons at CMS and ATLAS}

\author{ HENNING KIRSCHENMANN, \begin{textit} on behalf of the ATLAS and CMS collaborations \end{textit}}

\address{Helsinki Institute of Physics, Gustaf H\"allstr\"omin katu 2,\\
00014 Helsinki, Finland}

\maketitle\abstracts{
Recent measurements performed by the ATLAS and CMS collaborations on the Run 2 dataset are testing QCD with unprecedented precision. The wealth of data, an ever-improving experimental understanding of jets and photons, and novel measurements of jet substructure enable an improved understanding.
}

\section{Introduction}
QCD is an extemely successful theory and has been tested by many experiments over the course of the last decades. However, there is a constant evolution in precision both by experimental measurements and theory predictions. Recent measurements performed using data from $pp$ collisions collected by the ATLAS \cite{Aad:2008zzm} and CMS \cite{CMS:2008zzk} experiments covered in this contribution are compared to state-of-the art predictions. These measurements are an important input to improving both the perturbative and non-perturbative aspects of the predictions. \let\thefootnote\relax\footnote{Copyright 2021 CERN for the benefit of the ATLAS and CMS Collaborations. CC-BY-4.0 license.}

In addition to cross section and event-shape measurements, there is a growing number of measurements of the substructure of jets, often devised in close connection with the theory community \cite{Dreyer:2018nbf,Gras:2017jty}, that provide new insight into several different scales of QCD simultaneously, from fixed order effects to perturbative shower and non-perturbative hadronization effects.

\section{Event observables}
\begin{figure}
\begin{minipage}{0.70\linewidth}
\centerline{\includegraphics[width=0.99\linewidth]{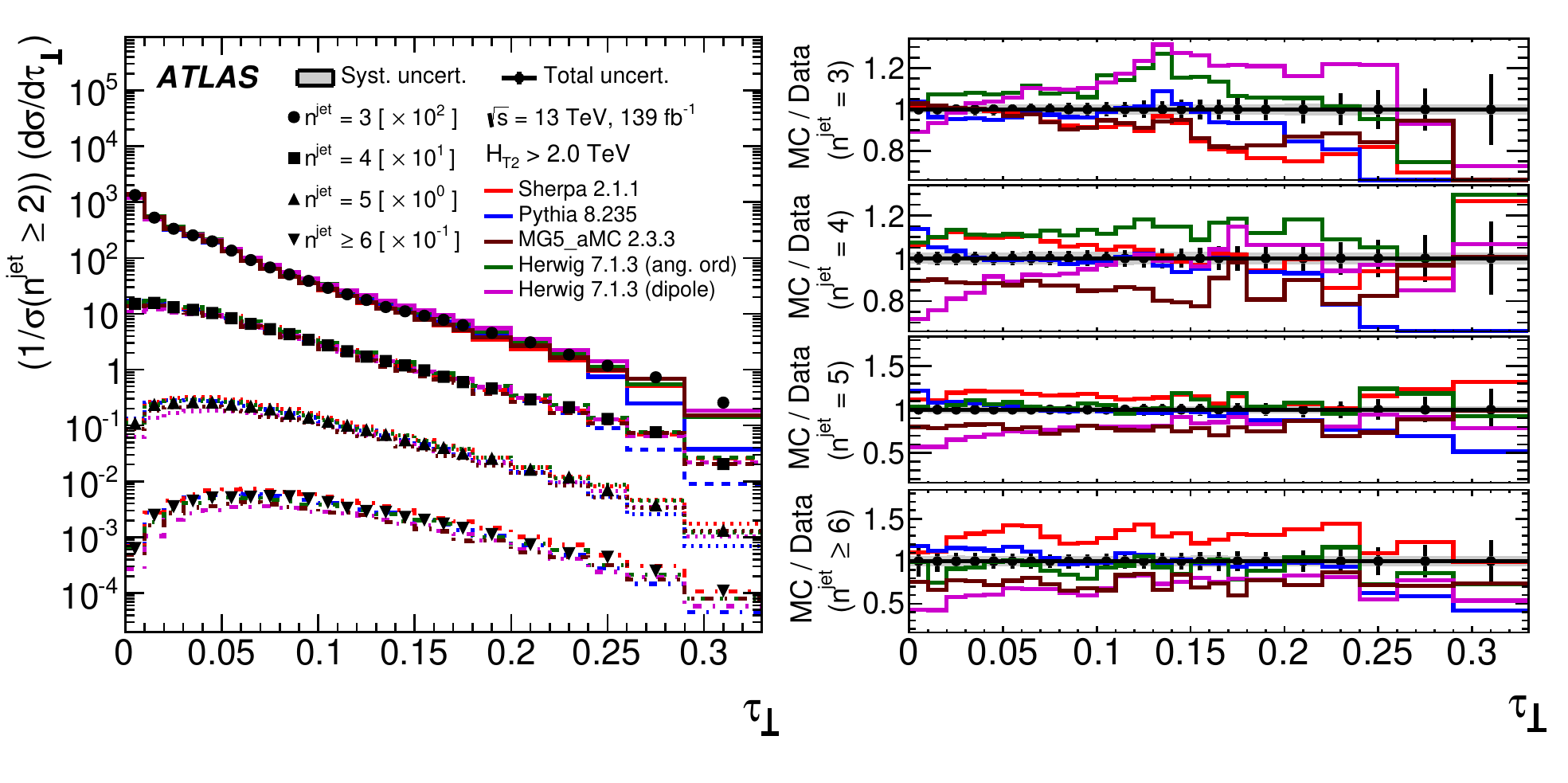}}
\end{minipage}
\hfill
\begin{minipage}{0.28\linewidth}
\centerline{\includegraphics[width=0.99\linewidth]{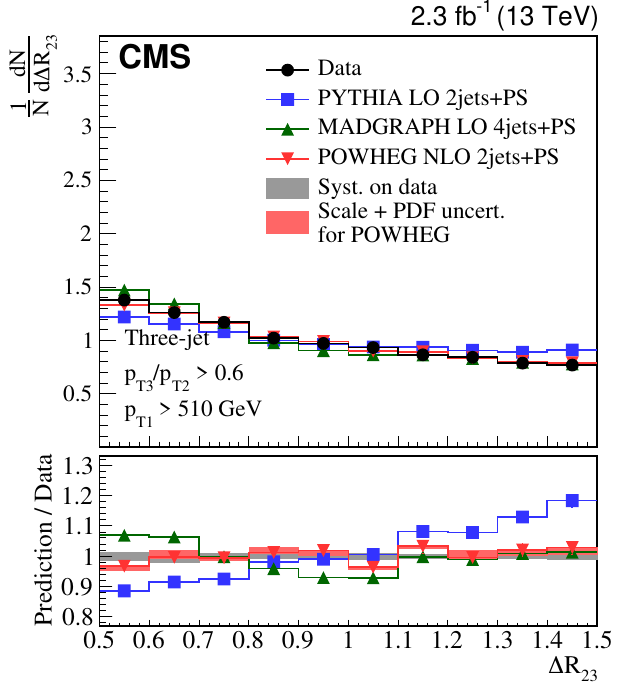}}
\end{minipage}
\caption[]{Left/Center: Comparison between data and MC simulation as a function of the transverse thrust for different jet multiplicites for $H_{T2}>2\,\textrm{TeV}$ (left) and the corresponding ratios between the MC and the data distributions.\cite{Aad:2020fch} Right: Three-jet events at $\sqrt{s}=13 \mathrm{TeV}$ and comparison to theoretical predictions: $\Delta R_{23}$ for hard $p_{\mathrm{T}}$ radiation $\left(p_{\mathrm{T} 3} / p_{\mathrm{T} 2}>0.6\right)$.\cite{Sirunyan:2021lwi}}
\label{fig:EventShapes}
\end{figure}

Event shapes\cite{Banfi:2004nk,Banfi:2010xy} are a proxy for energy flow in mutiljet final states and have been used in order to probe fundamental properties of QCD, tune Monte Carlo (MC) models, and in searches for physics beyond the Standard Model (SM). Six event-shape observables ($T_{\perp}$, $T_{m}$, $S$, $A$, $C$, and $D$) have been studied at different energy scales, given by the scalar sum of the two leading jets $H_{T2}$, in a recent measurement by ATLAS performed on the full Run 2 dataset \cite{Aad:2020fch}. 
The measurements are performed binned in jet multiplicity and $H_{T2}$. Final results are presented normalised to the inclusive cross section of events with at least two jets and unfolded to the particle level. The distributions for $\tau_{\perp}$, the complement of $T_{\perp}$, are shown in Figure~\ref{fig:EventShapes}. A general trend in most observables is that shape discrepancies are pronounced at low jet multiplicities, while at higher jet multiplicities the shapes are fairly well described albeit the normalization not agreeing between the predictions.
%
%

Main interest of a recently published measurement by CMS\cite{Sirunyan:2021lwi} related to angular and momentum distributions in multijet production at 8 and 13\,TeV is to understand the regimes of validity of the parton shower (PS) and matrix element approaches. Three jet or Z+2jet topologies, where the Z boson is treated equivalently to the leading jet, are analysed: The events are split into categories of interest by firstly cutting on the transverse momentum ratio to distinguish the soft from hard radiation regime and by secondly cutting on the angular separation to define a collinear and a large-angle radiation regime. The largest disagreements are observed in the collinear region, while the soft region is well described by the parton shower approach and large-angle and hard radiation regions are best described by multi-leg matrix element predictions. As an example, the unfolded distribution of the angular separation $\Delta R_{23}$  in the hard radiation region $\left(p_{\mathrm{T} 3} / p_{\mathrm{T} 2}>0.6\right)$ is shown in Figure~\ref{fig:EventShapes}.

\section{Inclusive jet, diphoton, and Z/$\gamma$+jet cross section measurements}
\begin{figure}
\begin{minipage}{0.38\linewidth}
\centerline{\includegraphics[width=0.99\linewidth]{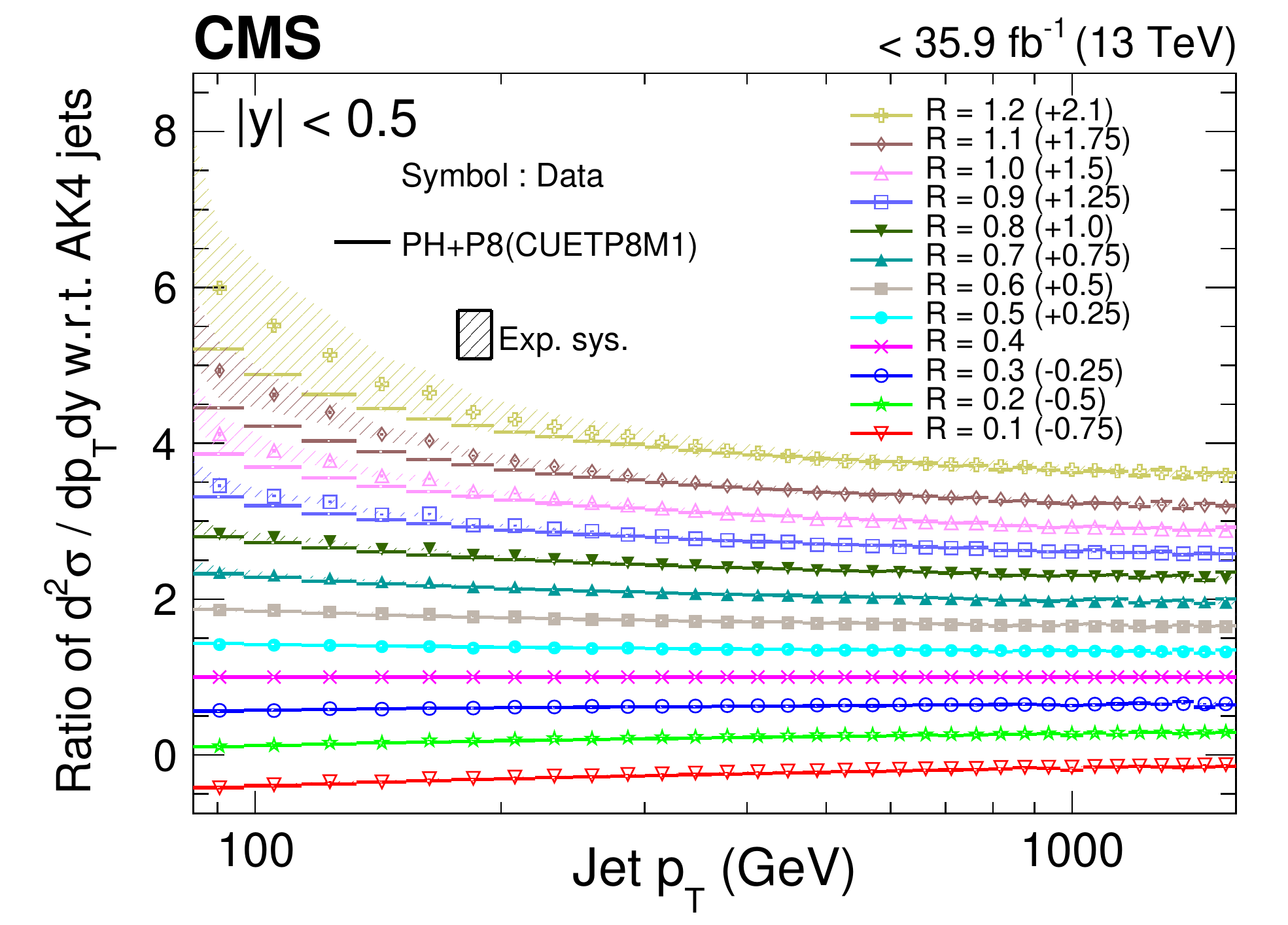}}
\end{minipage}
\hfill
\begin{minipage}{0.30\linewidth}
\centerline{\includegraphics[width=0.99\linewidth]{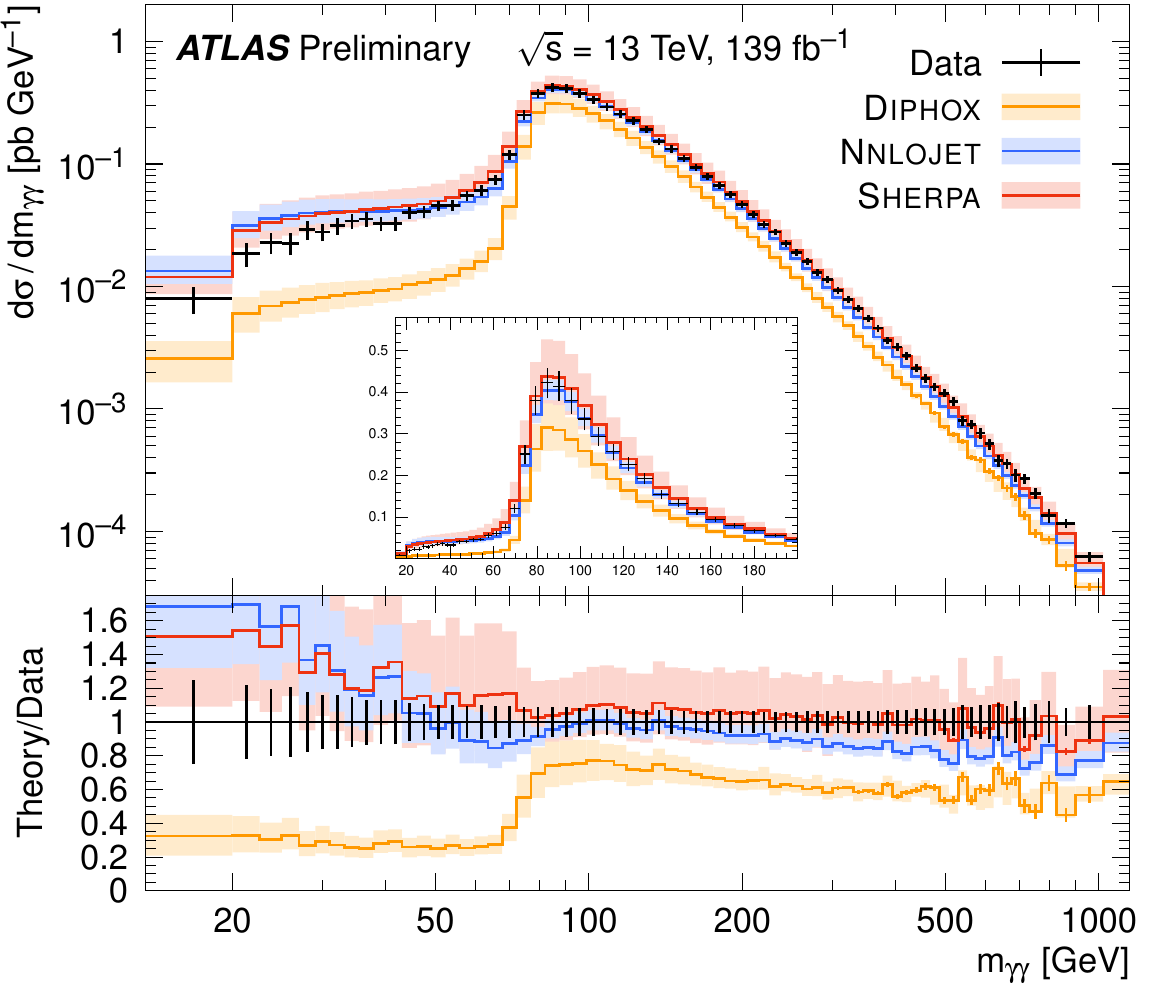}}
\end{minipage}
\hfill
\begin{minipage}{0.29\linewidth}
\centerline{\includegraphics[width=0.99\linewidth]{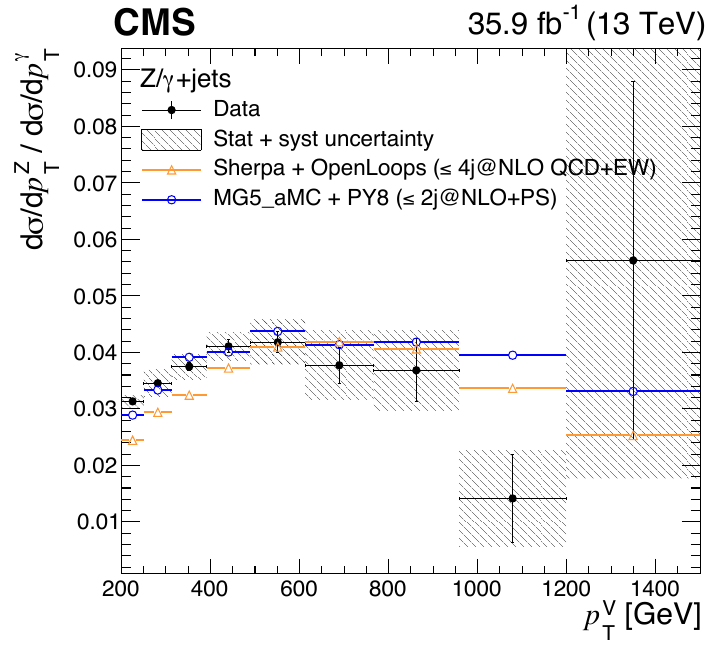}}
\end{minipage}
\caption[]{Left: Comparison of the ratio of the differential cross sections of jets of different sizes with respect to that of AK4 jets from data and from NLO predictions using POWHEG+PYTHIA (CUETP8M1 tune) in the region $|y| < 0.5$.\cite{Sirunyan:2020uoj} Middle: Differential diphoton cross section as a function of $m_{\gamma\gamma}$ compared to Diphox NLO, Nnlojet NNLO and Sherpa MEPS@NLO predictions.\cite{ATLAS:2020qqv} Right: Differential cross section ratio of Z+jets to $\gamma$+jets as a function of the vector boson (V) \pt compared with MADGRAPH5 aMC@NLO and SHERPA + OPENLOOPS predictions.\cite{Sirunyan:2021axe} }
\label{fig:CrossSections}
\end{figure}

In order to study the different components of the jet formation process in QCD, a scan through different distance parameters R of medium to high $p_{\mathrm{T}}$ jets has been performed by CMS on 2016 data.\cite{Sirunyan:2020uoj} The main observable is $(\mathrm{d} \sigma / \mathrm{d} y) /(\mathrm{d} \sigma / \mathrm{d} y$ of $\mathrm{AK} 4$ jets), the double-differential inclusive jet cross section ratio. Results are unfolded and then compared to leading order (LO) and higher order (NLO, NNLO) predictions with different PS and with and without non-perturbative corrections. The best agreement is achieved by NLO+PS predictions over a wide range of distance parameter values. For small values of R we expect high sensitivity to PS effects and hadronization while for larger values of R the modelling of underlying event activity is studied in detail. In Figure~\ref{fig:CrossSections}, the main prediction (NLO using POWHEG+PYTHIA) describes the data  well at high \pt, with growing disagreement at low \pt\, for large values of R, hinting at suboptimal underlying event modelling.

A recent measurement of the diphoton inclusive cross section on the full Run 2 data by ATLAS \cite{ATLAS:2020qqv} provides differential distributions in 8 observables, covering transverse momenta, invariant masses and angular observables. The main experimental challenge is the background from non-prompt photons in jet events, around a third of the sample are $\gamma$+jet or jet-jet events with jets misidentified as photons. In Figure~\ref{fig:CrossSections}, the invariant mass of the diphoton system is shown: The mass region up to 1\,TeV is covered and a fine binning from lowest to highest masses exploits the full detector resolution. NLO and NNLO predictions generally describe the data well, except for the low invariant mass region, which is suppressed due to event selections and mostly populated by challenging to model $\gamma\gamma$+multijet events.

The goal of a recent CMS measurement \cite{Sirunyan:2021axe} is to provide the differential cross section separately for Z and $\gamma$+jet, and as a ratio, presented as a function of the vector boson \pt. The Z+jet cross section is also provided as a function of the distance between the Z boson and the closest jet in order to explicitly study the collinear Z emission. The Z/$\gamma$ ratio is sensitive to higher order electroweak corrections in particular in the high \pt\,range and plays an important role for data-driven background estimations in searches for physics beyond the SM. Figure~\ref{fig:CrossSections} shows the \pt\, dependence of the Z/$\gamma$ ratio: The data are generally in agreement with state-of-the-art NLO predictions.

\section{Jet substructure measurements}
\begin{figure}
\begin{minipage}{0.43\linewidth}
\centerline{\includegraphics[width=0.99\linewidth]{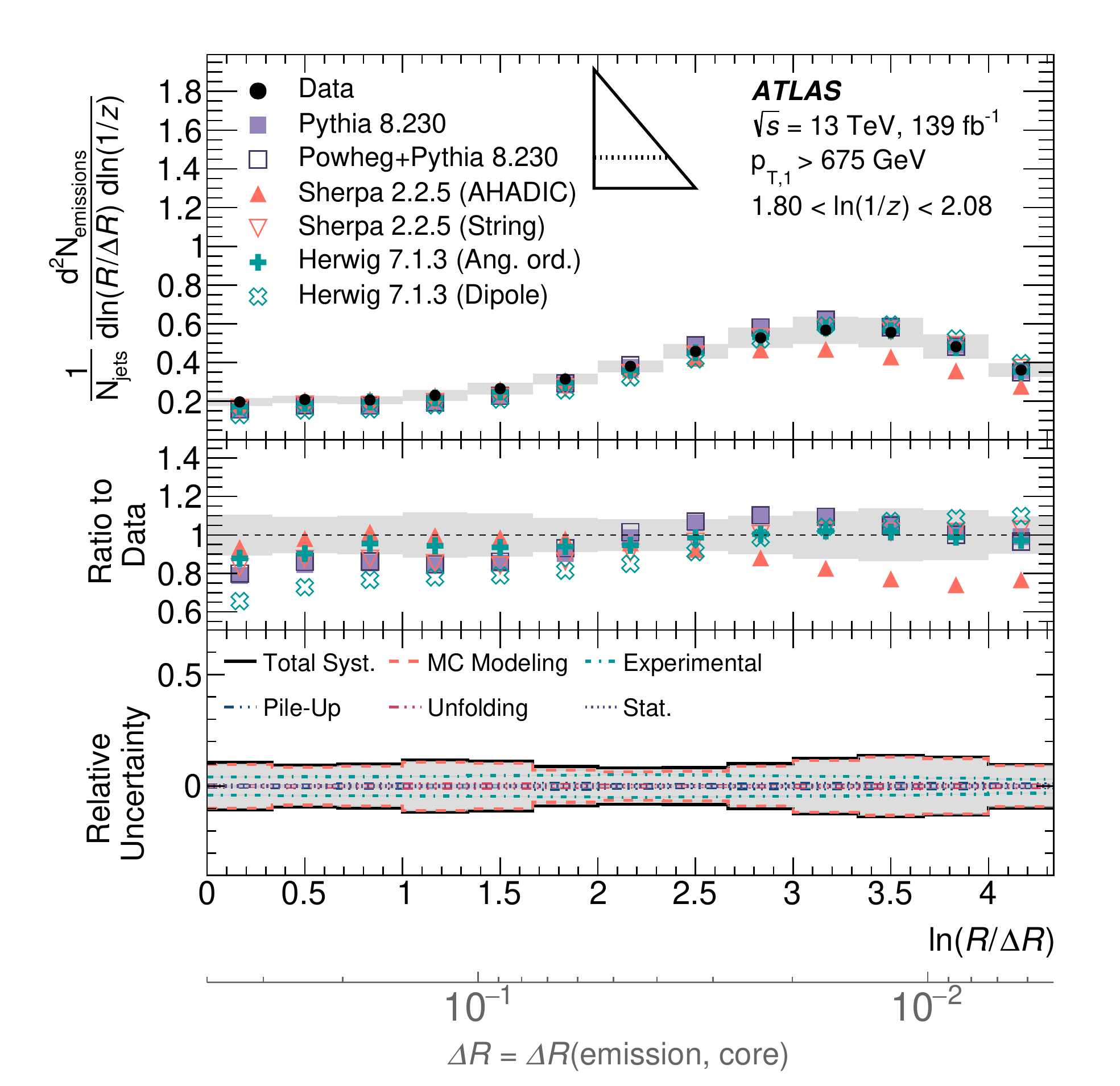}}
\end{minipage}
\hfill
\begin{minipage}{0.55\linewidth}
\centerline{\includegraphics[width=0.99\linewidth]{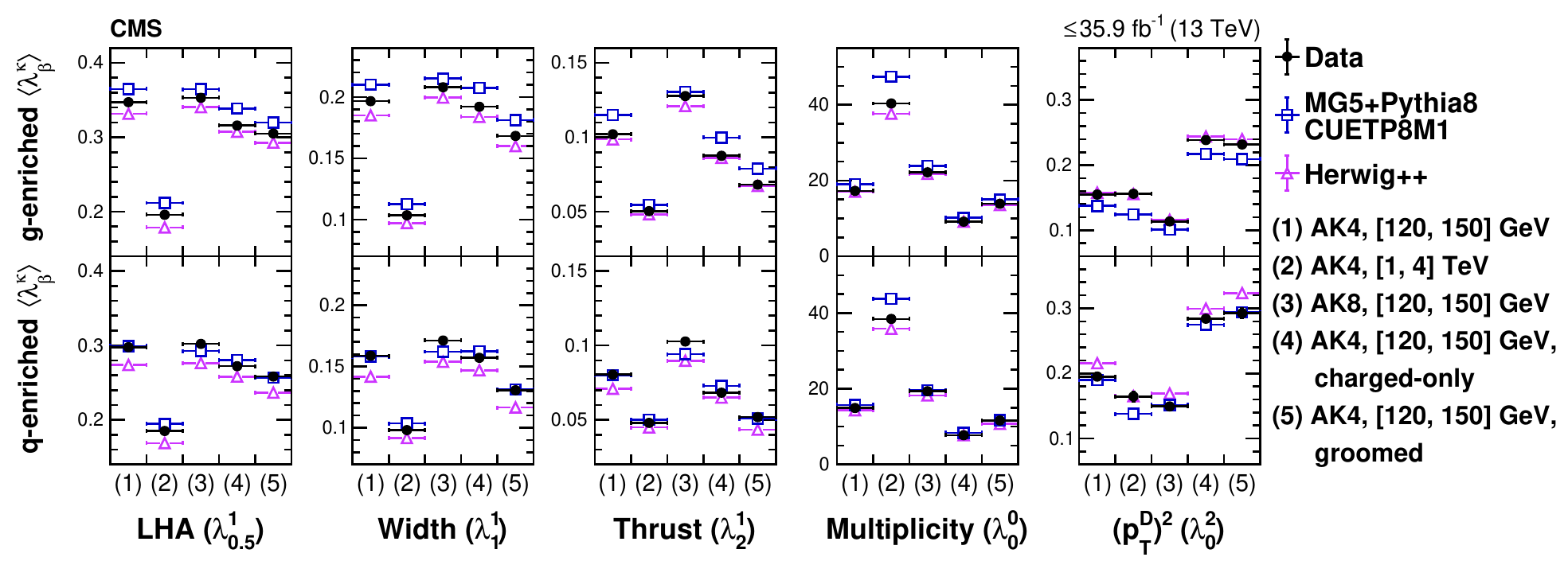}}
\centerline{\includegraphics[width=0.99\linewidth]{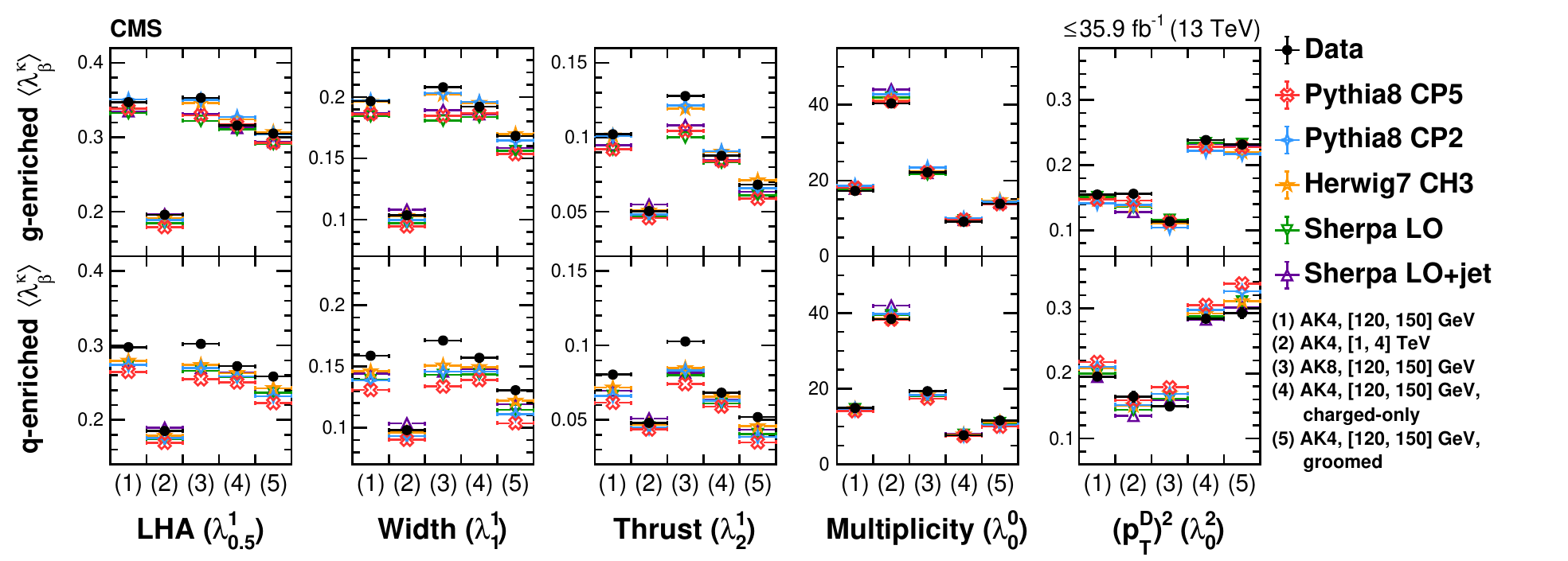}}
\end{minipage}
\caption[]{Left: Representative slice of the Lund Jet Plane ($1.80<\ln (1 / z)<2.08$). Various particle-level predictions are compared to unfolded data.\cite{Aad:2020zcn} Right: The means of various substructure observables are shown in quark/gluon-enriched regions for a representative selection of \pt, distance parameter, and grooming configurations and compared to various particle-level predictions.\cite{CMS:2021vsp} }
\label{fig:Substructure}
\end{figure}

A phenomological proposal \cite{Dreyer:2018nbf} introduces the concept of the ``Lund Jet Plane'' (LJP) as a novel way to look beyond single observables, giving a general handle on many aspects of QCD by probing the entire jet clustering history. In this picture, a jet is approximated as a series of soft emissions around a hard core, where the hard core represents the originating quark or gluon. The LJP is then span up by keeping track of the emissions via $z$, the relative momentum of emission with respect to the jet core, and $\Delta R$, which is the angle of emissions relative to the jet core, factorizing perturbative and non-perturbative effects, UE, and MPI. The measurement published by ATLAS \cite{Aad:2020zcn} analyses high \pt\,jets and only considers charged particles, which have naturally a good correspondence to the particle level due to the good tracking capabilities. In order to facilitate comparisons of predictions to the data, slices of the LJP are used to summarize the results, chosen such that they are susceptible to various effects. An example of such a slice is shown in Figure~\ref{fig:Substructure}, being particularly sensitive to the modelling of the parton shower at low values of $ln(R/\Delta R)$ and to hadronization effects at high values of values of $ln(R/\Delta R)$. The predictions are specifically chosen to contrast LO vs. NLO and different hadronization/parton shower models. An analytical prediction\cite{Lifson:2020gua} of the primary LJP density agrees well with the experimental measurements in regions where non-perturbative effects are small.

The latest addition to the jet substructure measurements, made public during the conference, is a systematic study by CMS \cite{CMS:2021vsp} to understand the interplay of soft and hard physics in quark and gluon jets. Jet substructre observables are an important input to jet tagging, but while quark jets have been charted in detail at LEP, gluon jets are less well understood overall. Inspired by a phenomenological proposal \cite{Gras:2017jty}, the measurement provides unfolded particle-level distributions for a set of generalized angularities. These are determined in many configurations: the calculation of the angularities is done using charged-only or charged+neutral deposits, for two different jet radii (AK4/AK8 jets), groomed/ungroomed, and in signal regions with different compositions of quark/gluon jets (dijet and Z+jet topologies). 
The wealth of distributions is also provided in a condensed way, focusing on the mean of the substructure observables for a representative set of configurations across all dimensions under study. Two of these summaries are shown in Figure~\ref{fig:Substructure}, comparing a wide set of predictions to the unfolded data. The results suggest that more recent generators under study improve the modelling of gluon jets at the cost of poorer modelling of quark jets. Looking forward, this measurement enables a detailed analysis of the assumptions and approximations made in predictions of jet substructure, ultimately enabling improved predictions for future measurements and searches relying on jet substructure

\section{Conclusions}
The latest results from the ATLAS and CMS Collaborations on QCD with jets and photons are presented. A wealth of new results is becoming available with more data analysed, up to the full Run 2 dataset. There are also new types of measurements being pursued, most notably on jet substructure, promising a rich set of future measurements with ever-improving precision.



\end{document}